\documentclass[11pt]{article}
\usepackage[left=2.5cm,top=2.5cm,right=2cm,bottom=2.5cm]{geometry}
\usepackage{amsmath, amssymb}

\usepackage{graphicx,subfigure,epsfig,epsf,color}
\usepackage[utf8x]{inputenc}
\usepackage{slashed}
\usepackage{physics}
\usepackage{feynmf}
\usepackage{graphicx}
\usepackage{graphics}
\usepackage{epstopdf}
\usepackage{subfigure}
\usepackage{amsfonts}
\usepackage{sectsty}
\usepackage{cite}
\usepackage{pdflscape}
\begin{document}
 \date{}
\title{Observational Constraints on the $f(\phi,T)$ gravity theory}
\maketitle
\begin{center}
\author{Ashmita}\footnote{p20190008@goa.bits-pilani.ac.in},~Payel Sarkar\footnote{p20170444@goa.bits-pilani.ac.in},~Prasanta Kumar Das\footnote{pdas@goa.bits-pilani.ac.in} \\
\end{center}
 
\begin{center}
 Birla Institute of Technology and Science-Pilani, K. K. Birla Goa campus, NH-17B, Zuarinagar, Goa-403726, India
\end{center}
\vspace*{0.25in}

\abstract{We investigate inflation in modified gravity framework by introducing a direct coupling term between a scalar field $\phi$  and the trace of the energy momentum tensor $T$ as $f(\phi,T) = 2 \phi( \kappa^{1/2} \alpha T + \kappa^{5/2} \beta T^2) $ to the Einstein-Hilbert action. We consider a class of inflaton potentials (i) $V_0 \phi^p e^{-\lambda\phi}$, (ii)  $V_0\frac{ \lambda \phi^p}{1+\lambda\phi^p}$ and investigate the sensitivity of the modified gravity  parameters $\alpha$ and $\beta$ on the inflaton dynamics. 
We derive the potential slow-roll parameters, scalar spectral index $n_s$, and tensor-to-scalar ratio $r$ in the above $f(\phi,T)$ gravity theory and analyze the following three choices of modified gravity parameters~(i)  Case I:~ $\alpha \neq 0, ~\beta=0$ i.e. neglecting higher order terms,  (ii) Case II:~ $\alpha=0$, $\beta \neq 0$~ and do the analysis for $T^2$ term, (iii) Case III:~ $\alpha \neq 0$ and $\beta \neq 0$ i.e. keeping all terms. For a range of potential parameters, we obtain constraints on $\alpha$ and $\beta$ in each of the above three cases using the WMAP and the PLANCK data.}
\maketitle
\section{Introduction}
Several research has been conducted over the past few decades to describe the evolution of the Universe, both on a theoretical and observational level. Recent observational results from the redshift of type Ia supernova\cite{Reiss}, Cosmic Microwave Background (CMB) \cite{Kolb, Spergel} anisotropy from Planck\cite{PLANCK}, Wilkinson Microwave Anisotropy Probe (WMAP)\cite{WMAP}, Baryon Acoustic Oscillations (BAO)\cite{Anderson}, Large Scale Structures\cite{spergel1}, altogether point to an expanding Universe which seems to be in harmony with the standard cosmological theory, i.e. the famous $\Lambda$CDM\cite{Sahni, Ratra, caroll, Turner, Arturo} model within the General Relativity framework. But, the evidence for isotropic and homogenous Universe appears to be detrimental to the conventional Cosmological model due to the horizon, flatness \cite{Liddle}, fine tuning\cite{Sahni, Weinberg}, coincidence\cite{Sahni, Zlatev} problems, etc. However, cosmic inflation, a period of exponential expansion, in the early Universe offers quite a plausible solution to these problems \cite{Guth, Kolb, Linde, Starobinsky, Albrecht, Alberto, Sarkar}. This theoretical framework was developed by Guth \cite{Guth}, Linde\cite{Linde}, Starobinsky \cite{Starobinsky}, Albrecht and Steinhardt \cite{Albrecht} around forty years ago.  
The most straightforward method to study inflation is to consider a scalar field $\phi$ called Inflaton, which under the influence of a particular potential $V(\phi)$ along with the slow-roll approximation (where the kinetic terms are neglected with respect to the potential term) is used to examine the inflationary expansion of the universe \cite{Baumann, Kinney}. A host of inflaton potential has been extensively studied \cite{Gron, sarkar1} along with various cosmological parameters using density perturbation and power-spectrum and has been verified by the CMBR anisotropy measurement \cite{Martin, PLANCK}.\\
\noindent Though the Einstein's general relativity is accepted as the most suited model of gravity, it has few limitations in it. It does not explain the requirement of dark matter and dark energy, to fit with the cosmological data which has been regarded as one of the primary drivers behind research into alternate theories of Einstein's gravity. In this approach, the Einstein-Hilbert action is modified by adding some polynomial function of Ricci scalar $R$ (i.e. $f(R)$ gravity)\cite{Nojiri,samanta, buchdahl,capozziello,clifton, Nojiri1, Nojiri3,Oikonomou1,Oikonomou2,Odintsov,Nojiri4, Odintsov2,Nojiri5, Nojiri6, Bhatti}, or some function of the Ricci scalar $R$ and/or the trace of the energy-momentum tensor $T$ ($f(R,T)$ gravity) \cite{Harko, Singh, Jamil, Houndjo, Myrzakulov, Ashmita, Bhattacharjee, Gamonal, Martin3, Sahoo2, Chen} or some Gauss-Bonnet function ($f(G)$ gravity) \cite{Nojiri:2005jg ,Nojiri2, Oikonomou3, Oikonomou4} etc. 
\noindent The first work of Inflation in modified gravity was done in \cite{Bhattacharjee} using a quadratic potential. The same type of analysis has been done in the literature using several potentials like
power-law and natural and hill-top potentials in modified gravity. Starobinsky-type potential can also give compatible results with observational data in $f(R,T)$ gravity\cite{Gamonal}. \\
\noindent In this paper, we consider a class of modified gravity theory considering a specific form of $f(\phi,T)$, discuss the slow-roll inflation in the context of this theory and obtain the limits on modified gravity parameters by comparing estimated values of CMB parameters with WMAP and PLANCK.
\noindent
We have proposed a modified gravity model of the type $f(\phi,T) = 2 \phi( \kappa^{1/2} \alpha T + \kappa^{5/2} \beta T^2) $ which is an extension to the normal Einstein gravity or we can say that it is an add on to the usual $f(R,T)=R+2\alpha \kappa T$ gravity where we have promoted $\alpha$ to a field by introducing $\phi$ coupled with $T$ and $T^2$ terms, identified the $\phi$ as an inflaton. 
A comprehensive study with the inflaton $\phi$ couples to $T$ to all order does not exist in the literature and the present work is the first step towards that direction. Although a work with $\phi$ coupled with $T$ exists in the literature \cite{chen2, Nisha}, but it was limited to the linear order, and we have extended that study by considering a higher order term $ \sim \phi~T^2$ in $f(\phi,T)$ and considers a host of potentials, different from those chosen by Zhang {\it et. al} \cite{chen2}. \\
\noindent Here, we have considered three different cases~(i)~Case I: $\alpha \neq 0$ and $\beta =0$ which leads to $f(\phi,T) = 2 \alpha \sqrt{\kappa} \phi T $,  
(ii)~Case II: $\alpha = 0$ and $\beta \neq 0$ leading to $f(\phi,T) = 2 \alpha \kappa^{5/2} \phi T^2 $,  and finally, 
(iii)~Case III: $\alpha \neq 0$ and $\beta \neq 0$  where both the $T$ and $T^2$ terms are present in $f(\phi,T)$. In each of the above three cases we study the slow-roll inflation and investigate what kind of constraints follow on the modified gravity parameters $\alpha, \beta$ from the WMAP and Planck data in the parameter space of the inflaton potential. Note that, in the limit, $\alpha \rightarrow 0$  and $\beta \rightarrow 0$, this reduces to normal Einstein gravity where the inflaton $\phi$ decays out i.e. its number density falls to zero. \\
The paper is organized as follows: In section 2, we obtain the Einstein Field equations in the modified $f(\phi, T)$ gravity and derive the slow-roll parameters in this modified gravity theory. 
In section 3, we discuss the inflationary scenario for two different potentials, and derive the cosmological parameters such as scalar spectral index $n_s$, tensor to scalar ratio $r$, tensor spectral index $n_T$, respectively. These cosmological parameters have been subject to constraints in the parameter space of inflaton potential $(\lambda,p)$ within the context of modified $f(\phi, T)$ gravity. In section 4, we analyze our results and compare those  with the PLANCK 2018\cite{PLANCK} and the WMAP\cite{WMAP} data.

\section{Field equations in $f(\phi,T)$ gravity:}
The action for the modified gravity $f(\phi, T)$ model with the scalar field $\phi$ coupled with the energy-momentum($T$) tensor can be written as,
 \begin{equation}
 \label{action}
     \mathcal{S}= \frac{1}{2}\int d^4x \sqrt{-g} ~ F(R,T, \phi) +\int d^4x\sqrt{-g}~\mathcal{L}_m (\phi, \partial_\mu \phi)
 \end{equation}
 where $F(R,T, \phi) = \frac{1}{\kappa} R + f(\phi, T)$, $R$ is the trace of the Ricci tensor $R_{\mu \nu}$, and $T$ is the trace of energy-momentum tensor $T_{\mu\nu}$ of the matter present in the Universe. The modified gravity term $f(\phi, T)$ is a simple polynomial function of $\phi$ and $T$ and we propose the following form  
 \begin{equation}
 f(\phi,T) = 2 \phi( \kappa^{1/2} \alpha T + \kappa^{5/2} \beta T^2). 
 \end{equation}
 $g$ is the determinant of the metric tensor $g_{\mu\nu}$ and G is the Newtonian constant of Gravitation. \footnote{We use the natural units, $c = \hbar = 1$, set $\kappa=8\pi G(=1/M^2_{Pl})=1$ and choose the metric signature $(+,-,-,-)$.} If the Universe during the inflation era is dominated by a single inflaton field $\phi(t)$, which contributes to the matter Lagrangian $\mathcal{L}_m$ given by,
 \begin{equation}
     \mathcal{L}_m(\phi, \partial_\mu \phi) = \frac{1}{2}g^{\mu\nu}\partial_{\mu}\phi\partial_{\nu}\phi-V(\phi) = \frac{1}{2} \dot{\phi}^2 - V(\phi)
 \end{equation}
 where in the last equality we have assumed the inflaton field is spatially homogeneous and it depends on $t$ only. 
 Varying the action Eq.~(\ref{action}) with respect to gravity $g^{\mu\nu}$, we get the modified Einstein equation as,
 \begin{equation}
     F_R(R,T, \phi) R_{\mu\nu}-\frac{1}{2}F(R,T, \phi) g_{\mu\nu}+\bigl(g_{\mu\nu}\Box-\nabla_{\mu}\nabla_{\nu} \bigr)F_R(R,T, \phi)=T_{\mu\nu}-F_T(R,T, \phi)T_{\mu\nu}-F_R(R,T, \phi)\Theta_{\mu\nu}
 \end{equation}
 Here $F_R(R,T,\phi)=\frac{\partial F(R,T,\phi)}{\partial R}$, $F_T(R,T,\phi)=\frac{\partial F(R,T,\phi)}{\partial T}$.  
The energy-momentum tensor for the inflaton $\phi(t)$ field  is, 
 \begin{equation}
 T_{\mu\nu}=-g_{\mu\nu}\mathcal{L}_m+2 \frac{\delta \mathcal{L}_m}{\delta g^{\mu\nu}} = \partial_\mu \phi \partial_\nu \phi - g_{\mu\nu} \Big(  \frac{1}{2} \dot{\phi}^2 - V(\phi)\Big),
 \end{equation}
the trace of the energy-momentum tensor $T (= g^{\mu\nu} T_{\mu\nu}) = -  \dot{\phi}^2 + 4 V(\phi)$ and $\Theta_{\mu\nu}=g^{\alpha\beta}\frac{\partial T_{\alpha\beta}}{\partial g^{\mu\nu}}$. Taking the inflaton field is spatially homogeneous, it takes the form of a perfect fluid, with $\mathcal{L}_m (= \frac{1}{2} \dot{\phi}^2 - V(\phi)) =-p $ which yields 
\begin{eqnarray}
\Theta_{\mu\nu}=g^{\alpha\beta}\frac{\partial T_{\alpha\beta}}{\partial g^{\mu\nu}}= -2T_{\mu\nu}-pg_{\mu\nu} = - 2 \partial_\mu \phi \partial_\nu \phi - 3 p  g_{\mu\nu}
\end{eqnarray}

The functional form of $f(\phi,T)$ considered here as 
$f(\phi,T) = 2 \phi( \alpha T + \beta T^2)$ (setting $\kappa = 1$). Accordingly, the Einstein equation takes the form as follows
 \begin{equation}
     R_{\mu\nu}-\frac{1}{2}R g_{\mu\nu}=T_{\mu\nu}^{eff}
 \end{equation}
where 
\begin{equation}
T_{\mu\nu}^{eff}=T_{\mu\nu}+2\phi \Bigl\{T_{\mu\nu}\bigl(\alpha+2\beta T \bigr)+p g_{\mu\nu}\bigl(\alpha+2\beta T \bigr)+\frac{1}{2}g_{\mu\nu} \bigl(\alpha T+\beta T^2 \bigr)\Bigr\}
\end{equation}
\noindent As mentioned earlier, assuming that the Universe is filled up with a single and homogeneous inflaton field, the effective energy-momentum tensor of the inflaton field will take a diagonal form and we can define the effective energy density $\rho_{eff}$ and pressure $p_{eff}$ as,
\begin{equation}
\label{rho}
    \rho_{eff} = T^{eff}_{00}=  \frac{1}{2}\dot{\phi}^2+V+2\phi\dot{\phi}^2 \bigl(\alpha-2\beta\dot{\phi}^2+8\beta V \bigr)+\phi \bigl(-\dot{\phi}^2+4V \bigr) \bigl(\alpha-\beta\dot{\phi}^2+4\beta V \bigr)
\end{equation}
 \begin{equation}
 \label{pressure}
      g_{ij} ~p_{eff} = T^{eff}_{ij} = \left[\frac{1}{2}\dot{\phi}^2-V-\phi \bigl(4V-\dot{\phi}^2 \bigr) \bigl(\alpha-\beta\dot{\phi}^2+4\beta V \bigr) \right] g_{ij}
 \end{equation}
  The trace of energy-momentum tensor will take the form
 \begin{equation}
     T=\rho_{eff}-3p_{eff}=-\dot{\phi}^2+4V+4\phi \bigl(-\dot{\phi}^2+4V \bigr) \bigl(\alpha-\beta\dot{\phi}^2+4\beta V \bigr)+2\phi\dot{\phi}^2 \bigl(\alpha-2\beta\dot{\phi}^2+8\beta V \bigr)
 \end{equation}
 The line element for the Friedman-Lemaitre-Robertson-Walker (FLRW) metric in spherical coordinates has the following form,
 \begin{equation}
     ds^2 = dt^2-a^2(t)\left[\frac{dr^2}{1-k r^2} + r^2 d\theta^2 + r^2 \sin^2 \theta d\phi^2  \right]
     \label{metric}
 \end{equation}
 where $k=0$ is for flat universe.
 For the background FLRW metric, the Friedman equations with effective density and pressure become
 \begin{equation}
 \label{friedman}
     3H^2=\rho_{eff}, ~-2\dot{H}-3H^2=p_{eff}
 \end{equation}
 The continuity equation for the effective energy density and pressure will be,
 \begin{equation}
 \begin{split}
 & \ddot{\phi}+3H\dot{\phi}\biggl\{1+2\phi\Bigl\{\alpha+2\beta \bigl(-\dot{\phi}^2+4V \bigr)\Bigr\}\biggr\}+V_{,\phi}+2\dot{\phi}^2\Bigl\{\alpha+2\beta \bigl(-\dot{\phi}^2+4V \bigr)\Bigr\}+4\phi\ddot{\phi}\Bigl\{\alpha+2\beta(-\dot{\phi}^2+\\
 & 4V)\Bigr\}+4\beta\phi\dot{\phi}^2\bigl(-2\ddot{\phi}+4V_{,\phi}\bigr)+\bigl(-\dot{\phi}^2+4V\bigr)\Bigl\{\alpha+\beta \bigl(-\dot{\phi}^2+4V \bigr)\Bigr\}+\alpha\phi\bigl(-2\ddot{\phi}+4V_{,\phi}\bigr)+2\beta\phi \bigl(-\dot{\phi}^2+\\
 & 4V \bigr)\bigl(-2\ddot{\phi}+4V_{,\phi} \bigr)=0
 \end{split}
 \end{equation}
 \subsection{Slow-roll parameters and CMB constraints:}
 We assume that the universe is filled with spatially homogeneous scalar field which is minimally coupled with the trace of the energy-momentum tensor. When the potential energy term prevails over the kinetic energy term i.e. $V(\phi) >>\frac{1}{2}\dot{\phi}^2 $, a condition known as slow-roll condition, we enter the inflationary phase. To study the inflation, we define the slow-roll parameters.
 The  Hubble slow-roll parameters are defined as
 \begin{equation}
     \epsilon_H=-\frac{\dot{H}}{H^2}, ~~
     \eta_H=-\frac{1}{H}\frac{\ddot{\phi}}{\dot{\phi}}
 \end{equation}
 The slow-roll approximation reads,
 \begin{equation}
 \label{condition}
     \dot{\phi}^2<< V, ~~ \ddot{\phi}<<3H\dot{\phi}, ~~ \dot{\phi}^2<<H \dot{\phi}
 \end{equation}
 Applying these conditions into Eq.~(\ref{friedman}) along with Eq.~(\ref{rho}) and Eq.~(\ref{pressure}), we get,
 \begin{equation}
 \label{H}
     3H^2 = V \bigl(1+4\alpha\phi+16\beta\phi V \bigr),
\end{equation}
\begin{equation}
\label{continuity}
    3H\dot{\phi} \Bigl\{1+2\phi \bigl(\alpha+8\beta V \bigr)\Bigr\}+\frac{dV}{d\phi}+4\alpha\phi V_{,\phi}+32\phi\beta V V_{,\phi}+4V \bigl(\alpha+4\beta V \bigr)=0
 \end{equation}
 In $f(\phi,T)$ gravity, we can find $\dot{H}$ from Eq.~(\ref{friedman}) as,
 \begin{equation}
 \label{dotH}
     \dot{H}=-\frac{1}{2}\bigl(\rho_{eff}+p_{eff} \bigr)=-\frac{1}{2}\dot{\phi}^2 \biggl[1+2\phi\Bigl\{\alpha+2\beta \bigl(-\dot{\phi}^2+4V \bigr) \Bigr\}\biggr]
 \end{equation}
 From Eq.~(\ref{H}) and Eq.~(\ref{dotH}) we can define the first potential slow-roll parameter as,
 \begin{equation}
 \epsilon_v = \frac{3\dot{\phi}^2}{2V} \biggl\{ \frac{1+2\alpha\phi+16\beta\phi V}{1+4\alpha\phi+16\beta\phi V}\biggr\} = \frac{\Bigl\{V_{,\phi}\bigl(1+4\alpha\phi \bigr)+4V \bigl(\alpha+4\beta V \bigr)+32\phi\beta  V V_{,\phi}\Bigr\}^2}{2V^2 \Bigl\{1+2\phi \bigl(\alpha+8\beta V \bigr)\Bigr\}\Bigl\{(1+4\phi \bigl(\alpha+4\beta V \bigr)\Bigr\}^2}
 \end{equation}
 and taking the derivative of Eq.~(\ref{continuity}), we can get the second potential slow-roll parameter as, 
 \begin{equation}
 \begin{split}
     \eta_v 
     & = \eta_H+\epsilon_H =\frac{1}{V \Bigl\{1+4\phi(\alpha+4\beta V)\Bigr\}\Bigl\{1+2\phi \bigl(\alpha+8\beta V \bigr)\Bigr\}}\biggl[V_{,\phi\phi} \bigl(1+4\alpha\phi \bigr)+4V_{,\phi} \bigl(2\alpha + 4 \beta V\bigr) +\\
     &\quad 16\beta V V_{,\phi}+ 32\beta \bigl(V V_{,\phi}+\phi V_{,\phi}^2+\phi V V_{,\phi\phi} \bigr)+\frac{\Bigl\{ V_{,\phi} \bigl(1+4\alpha\phi \bigr)+4V \bigl(\alpha+4\beta V \bigr)+32\phi\beta V V_{,\phi} \Bigr\} }{1+2\phi \bigl(\alpha+8\beta\phi \bigr)} \times \\
     &\quad \bigl(2\alpha+16\beta V+16\beta\phi V_{,\phi} \bigr)\biggr]
 \end{split}
 \end{equation}
 \noindent
 We see that in the limit $\alpha\rightarrow 0$ and $\beta\rightarrow 0$, the expression for the potential slow-roll parameters leads to normal Einstein gravity.  The scalar spectral index and tensor-to-scalar ratio can be expressed in terms of potential slow-roll parameters as follows,
 \begin{equation}\label{CMB}
     n_s=1-6\epsilon_v+2\eta_v, ~~  r=16\epsilon_v 
 \end{equation}
 Finally, the e-fold number $(N)$, can be defined as the amount of inflation needed to produce an isotropic and homogeneous Universe,
 \begin{equation}\label{efold}
     N=\int_{\phi_{in}}^{\phi_{final}}\frac{H}{\dot{\phi}}d\phi=\int_{\phi_{final}}^{\phi_{in}} \frac{V \Bigl\{1+4\phi \bigl(\alpha+4\beta V \bigr)\Bigr\} \Bigl\{1+2\phi \bigl(\alpha+8\beta V \bigr) \Bigr\}}{V_{,\phi} \bigl(1+4\alpha\phi \bigr)+4V \bigl(\alpha+4\beta V \bigr)+32\phi\beta VV_{,\phi}}d\phi
 \end{equation}

\section{Analysis of slow-roll inflation for different potentials:}
We consider the following three cases: (i) $\alpha\neq0$, and $\beta=0$, (ii) $\alpha = 0$, and $\beta \neq 0$ and (iii) $\alpha\neq0$, and $\beta \neq 0$. In each case, we study the slow-roll inflationary cosmology with two inflaton potentials
$ V=V_0 \phi^p e^{-\lambda\phi}$ and $V=V_0\frac{\lambda\phi^p}{1+\lambda\phi^p}$. We calculate the slow roll-parameters, e-fold, scalar spectral index and  and tensor-to-scalar ratio and obtain the range of $\alpha$ and $\beta$ using those quantities in the parameter space of the two potentials, which are consistent with WMAP and Planck data. 
\subsection{Case I: For $\alpha\neq0$, and $\beta=0$:} 
We start with $\beta = 0$  where $\phi$ couples linearly with $T$ and $f(\phi,T)$ takes the simplest form $f(\phi, T)= 2\sqrt{\kappa}~ \alpha \phi T$. We investigate the slow-roll inflationary expansion for two different inflaton potentials.
\subsubsection{Inflaton potential $V=V_0 \phi^p e^{-\lambda\phi}$}
The first inflaton potential, which is a combination of the power law and exponential term, is of the following form
\begin{equation}
    V=V_0 \phi^p e^{-\lambda\phi}
\end{equation}
where $V_0$ is a constant, p (power index) and $\lambda$ are the potential parameters. We find that for $\lambda=0$ reduces to ``chaotic potential $(\phi^p)$". On the other hand, $p=0$ leads to  exponential inflationary potentials with positive curvature discussed in \cite{Valentino}.
Under the slow-roll approximation, the slow-roll parameters in terms of potential parameters takes the form:
\begin{equation}
    \epsilon_v = \frac{1}{2(1+2 \alpha \phi)} \left[ \frac{4 \alpha}{1+4 \alpha \phi} + \frac{p-\lambda \phi}{\phi} \right]^2
\end{equation}

\begin{equation}\label{etav1}
\begin{split}
    \eta_v 
    &= \frac{1}{\phi^2 (1+2 \alpha \phi)^2 (1+4\alpha \phi)} \times \biggl[ p^2 \bigl(1+2 \alpha \phi\bigr) \bigl(1+ 4\alpha \phi \bigr) - p \bigl\{1+2 \lambda \phi \bigl(1+2 \alpha \phi \bigr) \bigl(1+ 4\alpha \phi\bigr)\bigl\} + \phi^2 \Bigl\{\lambda^2 + \\
    &\qquad 6 \alpha \lambda \bigl(-1+\lambda \phi \bigr) + 8 \alpha^2 \bigl\{-1+\lambda \phi \bigl(-1+\lambda \phi \bigr)\bigl\}\Bigr\} \biggr]
\end{split}
\end{equation}
\noindent The scalar spectral index($n_s$) and tensor-to-scalar ratio($r$) can be expressed in terms of potential parameters, modified gravity parameter(s), inflaton field $\phi$ as follows:
\begin{equation}
\begin{split}
    n_s 
    &= 1 - \frac{3}{1+2 \alpha \phi} \biggl[ \lambda - \frac{p}{\phi} -\frac{4 \alpha}{1+4 \alpha \phi} \biggr]^2 + \frac{2}{\phi^2 (1+2 \alpha \phi)^2 (1+4\alpha \phi)} \biggl[ p^2 \bigl(1+2 \alpha \phi\bigr) \bigl(1+ 4\alpha \phi \bigr) - p \bigl\{1+\\
    &\qquad 2 \lambda \phi \bigl(1+2 \alpha \phi \bigr) \bigl(1+ 4\alpha \phi\bigr)\bigl\} + \phi^2 \Bigl\{\lambda^2 + 6 \alpha \lambda \bigl(-1+\lambda \phi \bigr) + 8 \alpha^2 \bigl\{-1+\lambda \phi \bigl(-1+\lambda \phi \bigr)\bigl\}\Bigr\} \biggr]
    \end{split}
\end{equation}
and
\begin{equation}
    r = \frac{8}{1+2 \alpha \phi} \left[ \frac{4 \alpha }{1+4 \alpha \phi} + \frac{p-\lambda \phi}{\phi} \right]^2
\end{equation}

\subsubsection{Inflaton potential $V=V_0\frac{\lambda\phi^p}{1+\lambda\phi^p}$}

\begin{figure}[htb]
 \centerline {\psfig{file=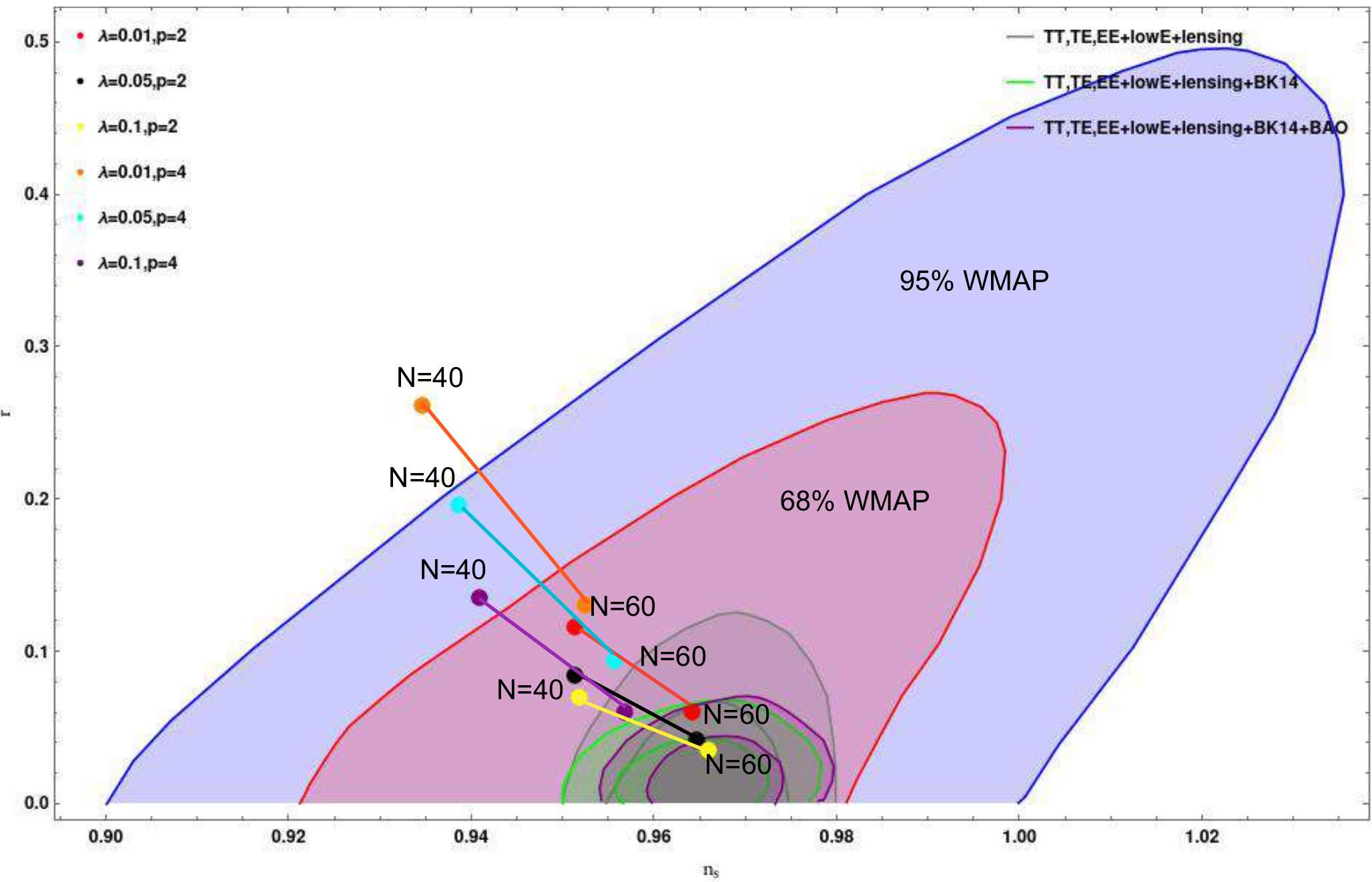,width=6.5cm}\psfig{file=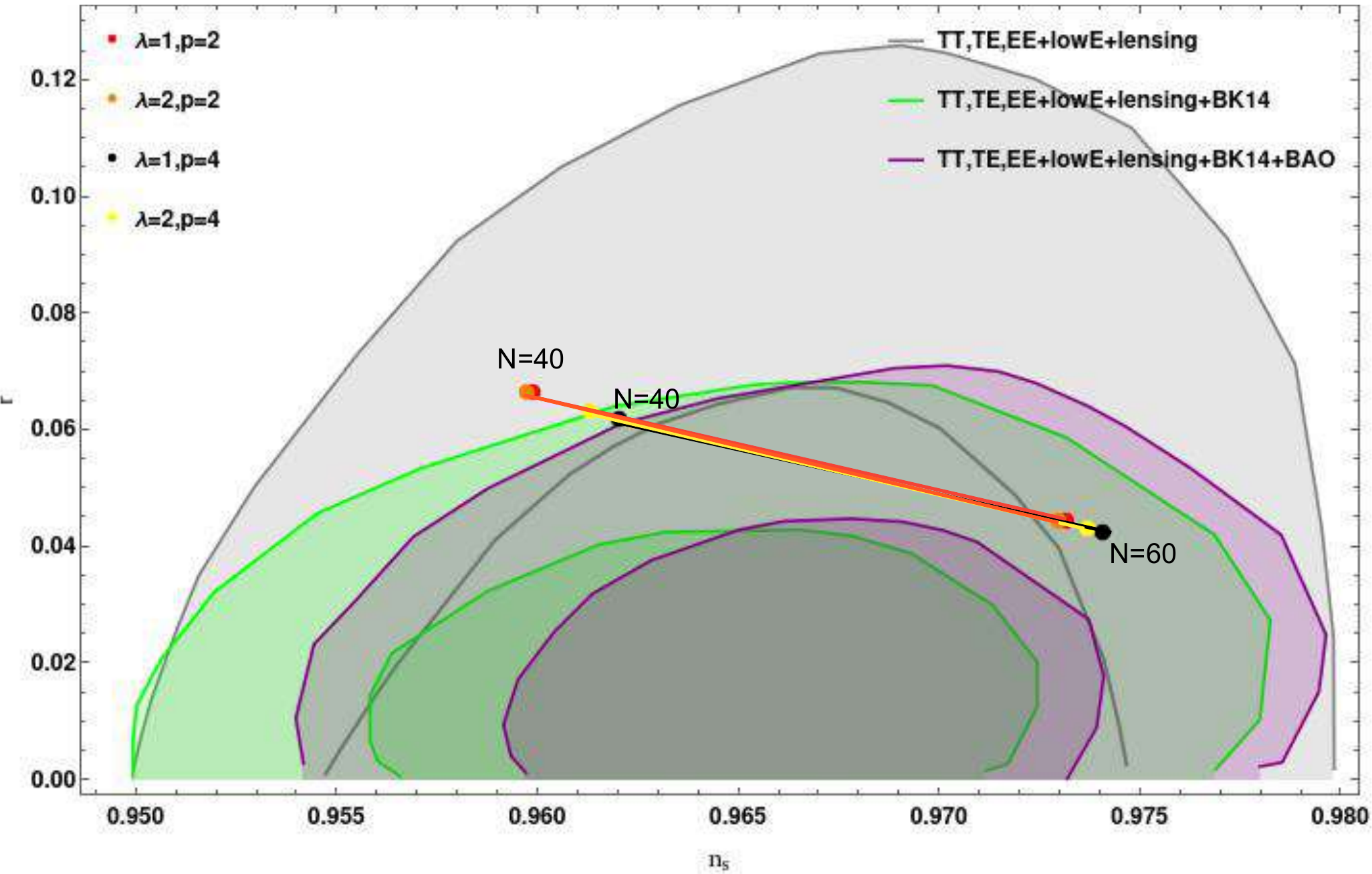,width=6.5cm}}
 \vspace*{3pt}
  \caption{(Color online) Constraints on $n_s$ and $r$ from CMB measurements of different potential. Shaded regions are allowed by WMAP measuremnts, PLANCK alone, PLANCK+BK15, PLANCK+BK15+BAO upto $68\%$ and $95\%$ Confidence Level. \label{Plot1}}
\end{figure}

Next, we consider the fractional potential for inflationary expansion,
\begin{equation}
    V=V_0\frac{\lambda\phi^p}{1+\lambda\phi^p}
\end{equation}
where $V_0$ is a constant, p and $\lambda$ are the potential parameters. We have taken $p = 2$, $4$ for this potential in this anaysis. For $p=2$, this takes the form of fractional potential which was first studied by Eshagli et al.\cite{Eshagli}. For $p=4$, this potential takes the form of Higgs inflation potential which was first studied by Maity\cite{Maity}. Both of these potentials have been studied in the minimal scenario in normal Einstein gravity and non-minimal coupled gravity\cite{sarkar1}. \\
We have incorporated the modified gravity terms in the Lagrangian to see the effect of coupling between T and $\phi$. 
Under the slow-roll approximation, the slow-roll parameters have the form:
\begin{equation}
    \epsilon_v = \frac{1}{2 \bigl(1+2 \alpha \phi \bigr)} \left[ \frac{4 \alpha}{1+4 \alpha \phi} + \frac{p}{\phi+\lambda \phi^{1+p}} \right]^2
\end{equation}

\begin{equation}
    \eta_v = -\frac{p + p\lambda \phi^p +p^2 \bigl( 1+6 \alpha \phi + 8 \alpha^2 \phi^2 \bigr) \bigl( -1 + \lambda \phi^p \bigr) +8 \alpha^2 \phi^2 \bigl( 1 + \lambda \phi^p \bigr)^2}{\phi^2 \bigl(1+2 \alpha \phi \bigr)^2 \bigl(1+4 \alpha \phi \bigr) \bigl(1+\lambda \phi^p \bigr)^2}
\end{equation}
The scalar spectral index and tensor-to-scalar ratio can be expressed in terms of potential parameters as follows:
\begin{equation}
\begin{split}
    n_s 
    &= 1 - 2 \times \biggl[\frac{p + p\lambda \phi^p +p^2 \bigl( 1+6 \alpha \phi + 8 \alpha^2 \phi^2 \bigr) \bigl( -1 + \lambda \phi^p \bigr) +8 \alpha^2 \phi^2 \bigl( 1 + \lambda \phi^p \bigr)^2}{\phi^2 \bigl(1+2 \alpha \phi \bigr)^2 \bigl(1+4 \alpha \phi \bigr) \bigl(1+\lambda \phi^p \bigr)^2} \biggr] - \frac{3}{1+2 \alpha \phi} \times \\
    & \qquad \biggl[ \frac{4 \alpha}{1+4 \alpha \phi}+ \frac{p}{\phi+\lambda \phi^{1+p}} \biggr]^2
    \end{split}
\end{equation}
and 
\begin{equation}
    r = \frac{8}{2(1+2 \alpha \phi)} \left[ \frac{4 \alpha}{1+4 \alpha \phi} + \frac{p}{\phi+\lambda \phi^{1+p}} \right]^2
\end{equation}
 We can determine the dependency of modified gravity parameter $\alpha$ in the context where the scalar field $\phi$ couples with the trace of energy-momentum tensor $T$. For the potential $V=V_0\phi^p e^{\lambda\phi}$, $\alpha$ lies between $[-0.00928,-0.00079],[-0.00912,-0.00625],[-0.00763,-0.00389]$ for $p=2$ and $\lambda=0.01,0.05,0.1$ respectively which brings it to a better agreement with observational PLANCK 2018 data for the spectral index parameter $n_s$ and the tensor-to-scalar ratio $r$ along with the e-fold number($N$) lying in the range $40<N<70$. Similarly, we find the range of $\alpha$ as $[-0.00711,-0.00669], [-0.00743,-0.00671]$ and $[-0.00736,-0.00588]$ for $p=4,\lambda=0.01,0.05,0.1$ respectively, which gives $n_s$ within $2\sigma$ limit of PLANCK 2018 data. Similarly, for the potential $V=V_0\frac{\lambda\phi^p}{1+\lambda\phi^p}$, the range of $\alpha$ is given by $ [0.24900, 0.54260]$, $[0.21520,0.72800]$ for $p=2,\lambda=1,2$ and $[0.26870,0.84410], [0.34070,0.97690]$ for $p=4,\lambda=1,2$ respectively. 
In Table .~(\ref{table:1a}) and Table.~(\ref{table:1b}), we have tabulated the values of cosmological parameters for a particular value of $\alpha$ (chosen from the given range) and $\phi$ for both of these two potentials.
\begin{table}[htb]
\centering
\small
\addtolength{\tabcolsep}{0.5pt}
\begin{tabular}{cccccccc}
\hline
Potential, & $V=V_0 \phi^{p} e^{-\lambda\phi}$, & $p = 2$ & & & & &  \\
Range of $\alpha$ & $\alpha$ & $\lambda$ & $\phi $ &  $\phi_f$ & N & $n_s$ & r \\ 

\hline
$-0.00928 < \alpha < -0.00079$ & -0.00754 & 0.01 & 14.48 & 1.38809 & 62 & 0.96481 & 0.05686 \\

$-0.00912 < \alpha < -0.00625$ & -0.00650 & 0.05 & 12.89 & 1.35256 & 60 & 0.96464 & 0.04193  \\

$-0.00763 < \alpha < -0.00389$ & -0.00111 & 0.1 & 10.58 & 1.31256 & 48 & 0.95624 & 0.04045 \\

\hline

Potential, & $V=V_0 \phi^{p} e^{-\lambda\phi}$, & $p = 4$ & & & & &  \\
Range of $\alpha$ & $\alpha$ & $\lambda$ & $\phi $ &  $\phi_f$ & N & $n_s$ & r \\ 
\hline

$-0.00711 < \alpha < -0.00669$ & -0.00685 & 0.01 & 23.01 & 2.80414 & 69 & 0.95537 & 0.09398 \\

$-0.00743 < \alpha < -0.00671$ & -0.00680 & 0.05 & 20 & 2.72642 & 61 & 0.95618 & 0.08971 \\

$-0.00736 < \alpha < -0.00588$ & -0.00650 & 0.1 & 17.48 & 2.63645 & 57 & 0.95525 & 0.06821 \\
\hline
\end{tabular}
\caption{\label{table:1a} For $V=V_0\phi^{p} e^{-\lambda\phi}$, the e-fold number $N$ and the spectral index parameters $n_s$ and $r$ calculated for a fixed value of $\phi$ and $\alpha$ are presented.}
\end{table}
\begin{table}[htb]
\centering
\addtolength{\tabcolsep}{1.5pt}
\small
\begin{tabular}{ccccccccc}
\hline
Potential, & $V=V_0\frac{\lambda\phi^p}{1+\lambda\phi^p}$, & $p = 2$ & & & & & & \\

Range of $\alpha$ & $\alpha$ & $\lambda$ & $\phi $ &  $\phi_f$ & N & $n_s$ & r \\ 
\hline
$0.24900< \alpha<0.54260$ & 0.4210 & 1 & 5.5 & 0.90641 & 61 & 0.97341 & 0.04387 \\
$0.21520< \alpha <0.72800$ & 0.57040 & 2 & 5.0 & 0.78771 & 61 & 0.97331 & 0.04386 \\
\hline
Potential, & $V=V_0\frac{\lambda\phi^p}{1+\lambda\phi^p}$, & $p = 4$ & & & & & \\
Range of $\alpha$ & $\alpha$ & $\lambda$ & $\phi $ &  $\phi_f$ & N & $n_s$ & r \\ 
\hline
$0.26870<\alpha<0.84410$& 0.66740 &1 & 4.7 & 1.08803 &58 &0.97335 & 0.04347 \\ 
$0.34070<\alpha<0.97690$ & 0.77660 & 2& 4.5 & 0.97349 & 59 & 0.97335 & 0.04352 \\
\hline
\end{tabular}

\caption{\label{table:1b} For $V=V_0\frac{\lambda\phi^p}{1+\lambda\phi^p}$, the e-fold number $N$ and the spectral index parameters $n_s$ and $r$, calculated for a fixed value of $\phi$ and $\alpha$ are presented.}
\end{table}
We can observe that all the values nicely match the data given by Planck 2018. Finally, in Fig.~{\ref{Plot1}} we have plotted the results of two potentials for $N=40$ and $60$. The blue and red shaded region corresponds to WMAP data up to $95\%$ and $68\%$ C.L whereas grey, green and purple shaded regions corresponds to PLANCK, PLANCK+BK15, PLANCK+BK15+BAO respectively.
 From left side of Fig.~(\ref{Plot1}), we can infer that for $p=2$ values agree with PLANCK data for $N=60$ whereas for $p=4$ the values match with WMAP data. All the values for $V=V_0\frac{\lambda\phi^p}{1+\lambda\phi^p}$ with $p=2,4$ and $\lambda=1,2$ lies within PLANCK data region as shown in right side of Fig.~(\ref{Plot1}).

\subsection{Case II: For $\alpha=0$ and $\beta\neq 0$:}
Next we are to investigate the impact of higher order term in the modified gravity parameter $f(\phi,T)$ i.e. $F(R,\phi,T)=R+2\beta\phi T^2$ on several cosmological parameters and their subsequent analysis. 
\subsubsection{Inflaton Potential $V=V_0\phi^p e^{-\lambda\phi}$}
As in the previous section, we derive the slow-roll parameters for this potential as:
\begin{equation}
    \epsilon_v = \frac{e^{\lambda \phi} \Bigl\{ e^{\lambda \phi} \bigl(-p + \lambda \phi \bigr) + 16 \beta \phi^{1+p} \bigl(-1-2p+2\lambda \phi \bigr) V_0 \Bigr\}^2 }{ 2 \phi^2 \bigl(e^{\lambda \phi} +16~\beta \phi^{1+p} ~V_0 \bigr)^3 }
\end{equation}
and
\begin{equation}
\begin{split}
    \eta_v 
    &= \frac{1}{\phi^2 \bigl(e^{\lambda \phi} +16~\beta \phi^{1+p}~ V_0 \bigr)^3} e^{\lambda \phi} \biggl[ e^{2 \lambda \phi} \Bigl\{p^2 + \lambda^2 \phi^2 -p \bigl(1+2 \lambda \phi \bigr)\Bigr\}+ 16 e^{\lambda \phi} \beta \phi^{1+p} \Bigl\{ 6 p^2 +p \bigl(2-12 \lambda \phi \bigr) \\
    & \quad + \lambda \phi \bigl(-5+6\lambda \phi \bigr) \Bigr\} V_0 + 256 \beta^2 \phi^{2+2p} \Bigl\{ 1+6p^2 -7\lambda \phi +6 \lambda^2 \phi^2 +p\bigl( 5-12\lambda \phi \bigr) \Bigr\} V_0^2 \biggr]
    \end{split}
\end{equation}
The spectral index $n_s$ and tensor to scalar ratio $r$ can be derived using Eq.~(\ref{CMB}) as,
\begin{equation}
    \begin{split}
        n_s 
        &= \frac{1}{\phi^2 \bigl(e^{\lambda \phi} +16~\beta \phi^{1+p}~ V_0 \bigr)^3} \biggl[ -e^{ 3\lambda \phi} \Bigl\{ p^2 + \bigl( -1+\lambda^2 \bigr)\phi^2 +2 p \bigl( 1 - \lambda \phi \bigr) \Bigr\} + 16 e^{2\lambda \phi}\beta \phi^{1+p} \Bigl\{ -2p +\phi \times \\
        &\quad \bigl( -4\lambda+3 \phi \bigr) \Bigr\} V_0 + 256 ~e^{\lambda \phi}\beta^2 \phi^{2+2p} \bigl( -1-2p-2\lambda \phi +3 \phi^2 \bigr) V_0^2 + 4096 \beta^3 \phi^{5+3p} V_0^3 \biggr]
     \end{split}
\end{equation}
and
\begin{equation}
    r = \frac{8~e^{\lambda \phi} \Bigl\{ e^{\lambda \phi} \bigl(-p + \lambda \phi \bigr) + 16 \beta \phi^{1+p} \bigl(-1-2p+2\lambda \phi \bigr) V_0 \Bigr\}^2 }{  \phi^2 \bigl(e^{\lambda \phi} +16~\beta \phi^{1+p} ~V_0 \bigr)^3 }
\end{equation}

\subsubsection{Inflaton Potential $V=V_0\frac{\lambda\phi^p}{1+\lambda\phi^p}$}
We follow the same procedure for the potential $V=V_0\frac{\lambda\phi^p}{1+\lambda\phi^p}$ as before. We calculate the slow-roll parameters as follows,
\begin{equation}
    \begin{split}
        \epsilon_V = \frac{\Bigl\{p+p\lambda\phi^p \bigl(1+32V_0\beta\phi \bigr)+16V_0\beta\phi^{1+p}\lambda \bigl(1+\lambda\phi^p \bigr)\Bigr\}^2}{2\phi^2 \bigl(1+\lambda\phi^p \bigr)\Bigl\{1+\lambda\phi^p \bigl(1+16V_0\beta\phi \bigr)\Bigr\}^3}
    \end{split}
\end{equation}
and 
\begin{equation}
    \begin{split}
        \eta_V
        &=\frac{1}{\phi^2 \bigl(1+\lambda\phi^p \bigr) \Bigl\{1+\lambda\phi^p \bigl(1+16V_0\beta\phi \bigr) \Bigr\}^3}\biggl[256V_0^2\beta^2\lambda^2\phi^{2+2p}(1+\lambda\phi^p)^2+p(1+\lambda\phi^p)\Bigl\{-1+\\
        &\quad 2\lambda\phi^p \bigl(-1+16V_0\beta\phi \bigr)+\lambda^2\phi^{2p}\bigl(-1+32V_0\beta\phi+1280V_0^2\beta^2\phi^2 \bigr)\Bigr\}+p^2\Bigl\{1+\lambda\phi^p \bigl(1+96V_0\beta\phi \bigr)-\\
        &\quad \lambda^3\phi^{3p}\bigl(1+48V_0\beta\phi+512V_0^2\beta^2\phi^2 \bigr)+\lambda^2\phi^{2p}\bigl(-1+48V_0\beta\phi+1536V_0^2\beta^2\phi^2 \bigr) \Bigr\}\biggr]
    \end{split}
\end{equation}
\begin{table}[htb]
\addtolength{\tabcolsep}{1.5pt}
\small
\begin{tabular}{ccccccccc}
\hline
Potential, & $V=V_0\phi^p e^{-\lambda\phi}$, & $p=2$ & & & & & & \\
Range of $\beta$ & $\beta$ & $\lambda$ & $\phi $ &  $\phi_f$ & N & $n_s$ & r \\ 
\hline
$1 \times 10^{-6}< \beta< 4 \times10^{-6}$ & $2\times 10^{-6}$ & 0 & 15 & 1.41434 & 55 & 0.97226 & 0.16864 \\
$5.8 \times 10^{-7}< \beta< 6.4 \times10^{-6}$ & $4\times 10^{-6}$ & 0.01 & 14.2 & 1.40452 & 51 & 0.97331 & 0.17254\\
$1.5 \times 10^{-5}< \beta< 6.9 \times10^{-5}$ & $4\times 10^{-5}$ & 0.1 & 11 & 1.32247 & 42 & 0.96963 & 0.08972\\
\hline
Potential, & $V=V_0\phi^p e^{-\lambda\phi}$, & $p=4$ & & & & & & \\

Range of $\beta$ & $\beta$ & $\lambda$ & $\phi $ &  $\phi_f$ & N & $n_s$ & r \\
\hline
$ 7.9 \times10^{-10}< \beta< 6.6 \times10^{-9}$ & $1 \times 10^{-9}$ & 0 & 21 & 2.82844 & 54 & 0.96502 & 0.35036 \\
$ 1.5 \times10^{-9}< \beta< 8.5 \times10^{-9}$ & $2.5 \times 10^{-9}$ & 0.01 & 20 & 2.80858 & 50 & 0.95649 & 0.32779 \\
$ 7 \times10^{-9}< \beta< 1 \times10^{-7}$ & $4 \times 10^{-8}$ & 0.1 & 17 & 2.64175 & 49 & 0.96478 & 0.18179 \\
\hline
\end{tabular}
\caption{\label{table:2a} For $V=V_0\phi^{p}e^{-\lambda\phi}$, the e-fold number $N$ and the spectral index parameters $n_s$, $r$ and $n_T$, calculated for a fixed value of $\phi$ and $\beta$ are presented.}
\end{table}
\noindent
The scalar spectral index and tensor to scalar ratio are obtained as,
\begin{equation}
\begin{split}
    n_s
    &= \frac{1}{\phi^2 \bigl(1+\lambda\phi^p \bigr)\Bigl\{1+\lambda\phi^p \bigl(1+16V_0\beta\phi \bigr)\Bigr\}^3}\Biggl[-2p \bigl(1+\lambda\phi^p \bigr)\Bigl\{1+2\lambda\phi^p \bigl(1+8\beta\phi V_0 \bigr)+\lambda^2\phi^{2p} \bigl(1+16V_0\beta\phi+\\
     &\quad 256V_0^2\beta^2\phi^2 \bigr)\Bigr\}-p^2 \biggl\{1+4\lambda\phi^p+\lambda^2\phi^{2p}\bigl(5+96V_0\beta\phi \bigr)+2\lambda^3\phi^{3p}\Bigl\{3+96V_0\beta\phi+256V_0^2\beta^2 \bigl(-1+3\phi^2 \bigr)\Bigr\}+\\
     &\quad \lambda^3\phi^{3p}\Bigl\{1+48V_0\beta\phi+4096V_0^3\beta^3\phi^3+256V_0^2\beta^2 \bigl(-1+3\phi^2 \bigr)\Bigr\} \biggr\} \Biggr]
    \end{split}
\end{equation}
and
\begin{equation}
    r=\frac{ 8 \Bigl\{p+p\lambda\phi^p \bigl(1+32V_0\beta\phi \bigr)+16V_0\beta\phi^{1+p}\lambda \bigl(1+\lambda\phi^p \bigr)\Bigr\}^2}{2\phi^2 \bigl(1+\lambda\phi^p \bigr)\Bigl\{1+\lambda\phi^p \bigl(1+16V_0\beta\phi \bigr)\Bigr\}^3}
\end{equation}

\noindent  We are to see now how the incorporation of the higher order terms in $f(\phi,T)$  affects the cosmological parameters and hence the constraints on $\beta$ in the potential parameter space. We find the range of $\beta$ for $p=2,\lambda=0,0.01,0.1$ are $[1\times 10^{-6}, 4\times 10^{-6}],[5.8\times 10^{-7},6.4\times 10^{-6}]$ and $[1.5\times 10^{-5},6.9\times 10^{-5}]$ whereas for $p=4,\lambda=0,0.01,0.1$ the values of $\beta$ lie within $[7.9\times 10^{-10},6.6\times 10^{-9}],[1.5\times 10^{-9},8.5\times 10^{-9}],[7\times 10^{-9},1\times 10^{-7}]$ which are compatible with the observational data. We have shown our results in  Table.~(\ref{table:2a}). 

\noindent Similarly, we have analyzed our another potential $V=V_0\frac{\lambda\phi^p}{1+\lambda\phi^p}$ and we have displayed the range of $\beta$ in table.~(\ref{table:2b}) 
\begin{table}[htb]
\addtolength{\tabcolsep}{1.5pt}
\small
\begin{tabular}{ccccccccc}
\hline
Potential, & $V=V_0\frac{\lambda\phi^p}{1+\lambda\phi^p}$, & $p = 2$ & & & & & & \\

Range of $\beta$ & $\beta$ & $\lambda$ & $\phi $ &  $\phi_f$ & N & $n_s$ & r \\ 
\hline
$10^{-8}< \beta<10^{-4}$ & $2\times 10^{-6}$ & 1 & 4.5 & 0.834946 & 56 & 0.972553 & 0.00351 \\
$10^{-7}< \beta <10^{-5}$ & $2\times 10^{-6}$ & 2 & 3.3 & 0.707113 & 32 & 0.95234 & 0.005676 \\
\hline
Potential, & $V=V_0\frac{\lambda\phi^p}{1+\lambda\phi^p}$, & $p = 4$ & & & & & \\
Range of $\beta$ & $\beta$ & $\lambda$ & $\phi $ &  $\phi_f$ & N & $n_s$ & r \\ 
\hline
$10^{-6}<\beta<10^{-4}$& $10^{-5}$ & 1 & 3.3 & 1.11390 &54 &0.96939 & 0.00085 \\ 
$10^{-8}<\beta<10^{-4}$ & $10^{-5}$ & 2& 2.95 & 0.98395 & 56 & 0.96993 & 0.00064 \\
\hline
\end{tabular}

\caption {\label{table:2b}For $V=V_0\frac{\lambda\phi^p}{1+\lambda\phi^p}$, the e-fold number $N$ and the spectral index parameters $n_s$, $r$ and $n_T$, calculated for a fixed value of $\phi$ and $\beta$ are presented.}
\end{table}
\begin{figure}[h]
 \centerline {\psfig{file=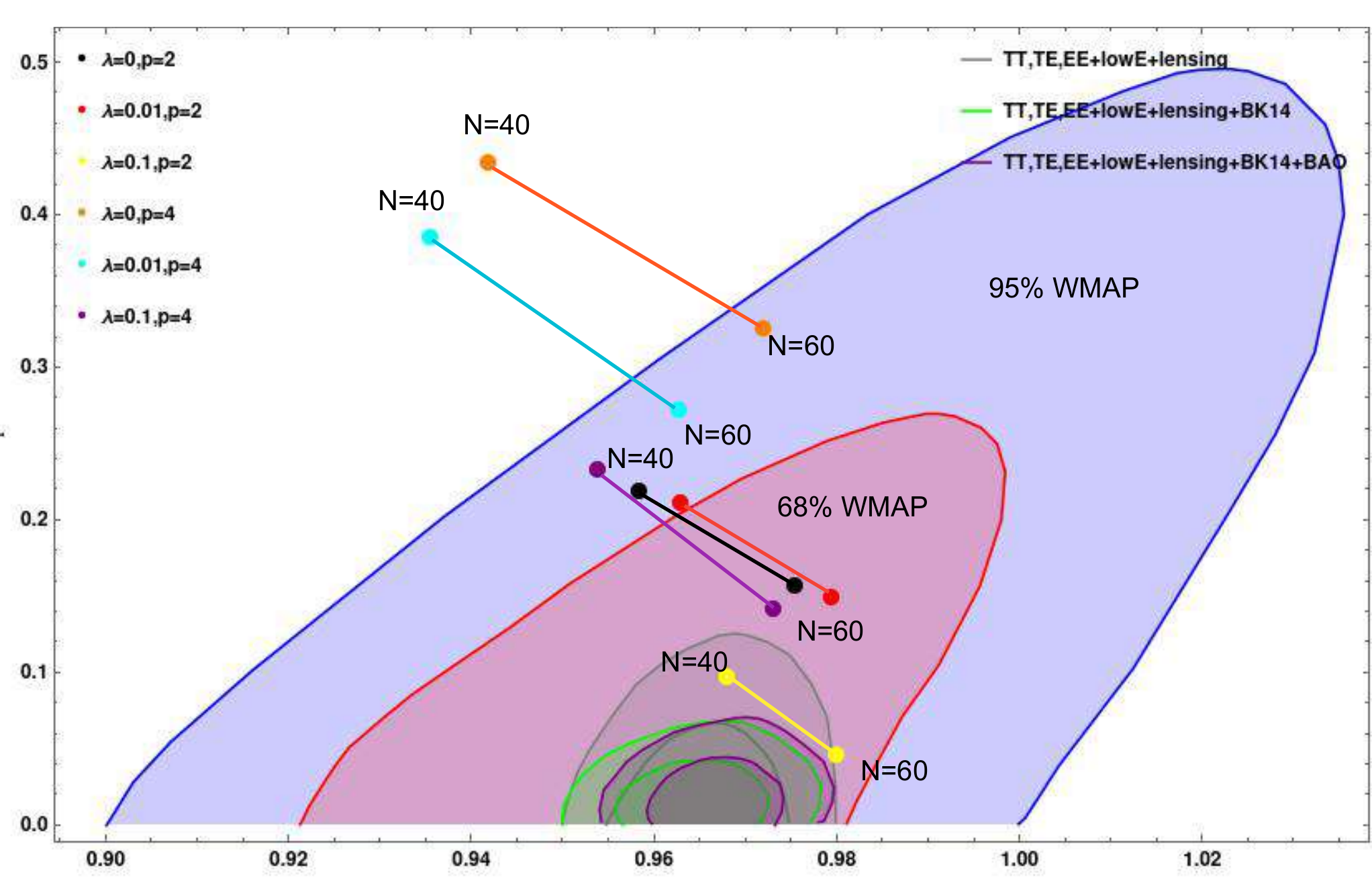,width=6.5cm}\psfig{file=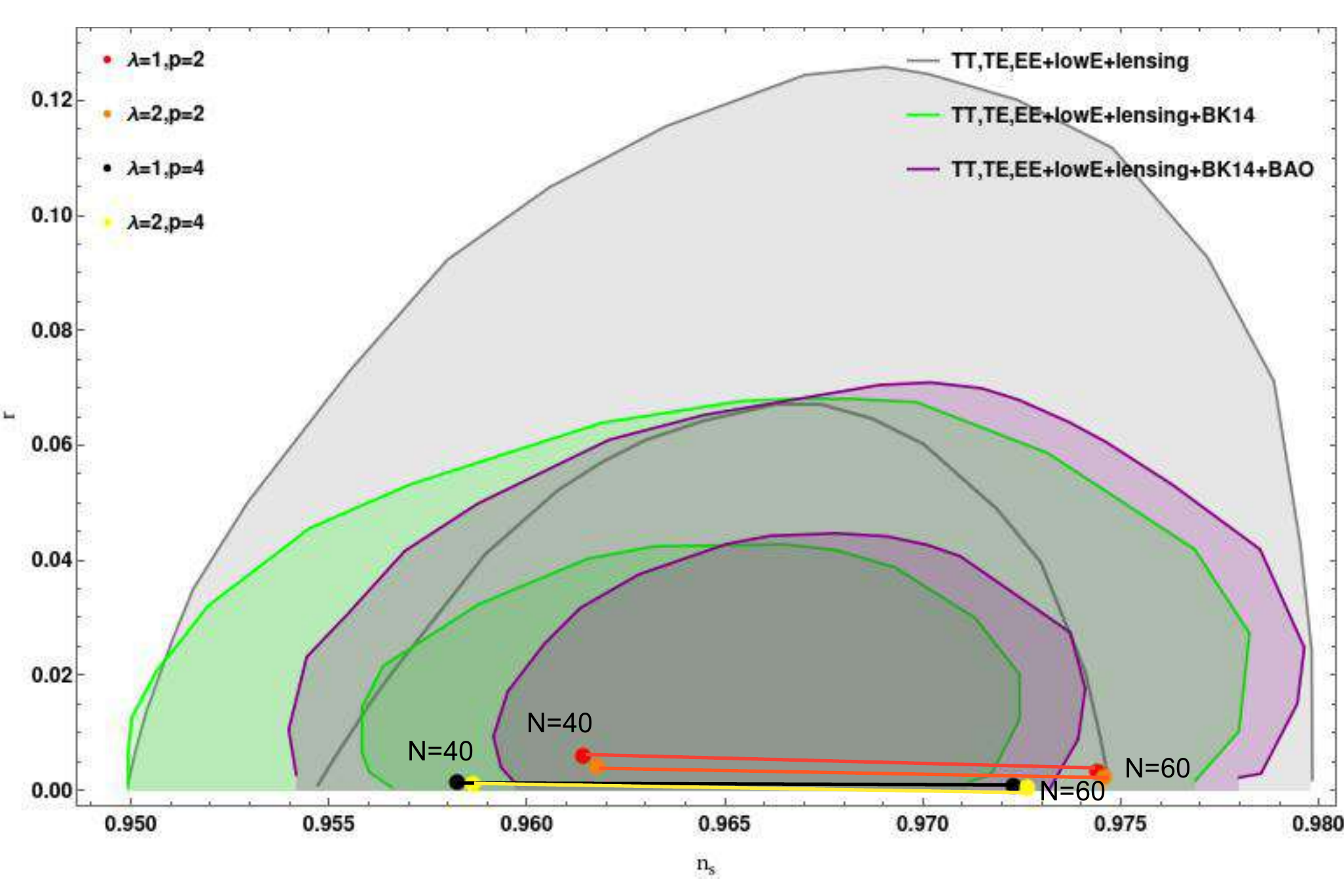,width=6.5cm}}
 \vspace*{3pt}
  \caption{(Color online) Constraints on $n_s$ and $r$ from CMB measurements of different potential. Shaded regions are allowed by WMAP measuremnts, PLANCK alone, PLANCK+BK15, PLANCK+BK15+BAO upto $68\%$ and $95\%$ Confidence Level. \label{Plot2}}
\end{figure}
In both Tables, the different cosmological parameter values are tabulated for particular value of $\phi$ and $\beta$(chosen from the range as shown). We clearly see that the inclusion of higher-order term produces large value of tensor-scalar ratio $r$ for $V_0\phi^p e^{\lambda\phi}$ and smaller values of $r$ for $V_0\frac{\lambda\phi^p}{1+\lambda\phi^p}$. Finally, in Fig. 2 we have plotted the results of two potentials for N = 40 and 60. From Fig.~(\ref{Plot2}) we can also realize that the potential $V_0\frac{\lambda\phi^p}{1+\lambda\phi^p}$ produces larger $r$ and hence is a better inflaton potential than $\phi^p e^{\lambda\phi}$ in modified gravity with $F(R,\phi,T)=R+2\beta\phi T^2$.
\subsection{Case III:~ Both $\alpha\neq0$ and $\beta\neq 0$:}
Finally, we consider the most general form i.e.  $f(\phi,T)=2 \phi( \kappa^{1/2} \alpha T + \kappa^{5/2} \beta T^2)$. We are to investigate how  the modifications in slow-roll parameters and spectral index parameters $n_s$ and $r$ results from this consideration. 
\subsubsection{Inflaton Potential $V=V_0\phi^p e^{-\lambda\phi}$}
After including both $T$ and $T^2$ terms, the slow-roll parameters become,
\begin{equation}
    \begin{split}
       \epsilon_V = \frac{e^{\lambda  \phi } \biggl\{e^{\lambda  \phi } \Bigl\{\phi  \bigl(4 \alpha  \lambda  \phi -4 \alpha +\lambda \bigr)-p \bigl(1 +4 \alpha  \phi \bigr) \Bigr\}+16 \beta V_0 \phi ^{p+1} \bigl(2 \lambda  \phi -2 p-1 \bigr) \biggr\}^2}{2 \Bigl\{ \bigl( 1 + 2 \alpha  \phi \bigr) e^{\lambda  \phi }+16 \beta  V_0 \phi ^{p+1}\Bigr\} \Bigl\{\phi \bigl( 1 +4 \alpha  \phi \bigr) e^{\lambda  \phi}+16 \beta V_0 \phi ^{p+2}\Bigr\}^2}
    \end{split}
\end{equation}
and 
\begin{equation}
    \begin{split}
        \eta_V
        & = \frac{1}{\Bigl\{ \bigl( 1+ 4 \alpha  \phi  \bigr) e^{\lambda  \phi }+16 \beta  V_0 \phi ^{p+1}\Bigr\} \Bigl\{\phi \bigl( 1+ 2 \alpha  \phi \bigr) e^{\lambda \phi }+16 \beta V_0 \phi^{p+2} \Bigr\}^2} \times  e^{\lambda \phi} \Biggl[ e^{2 \lambda  \phi } \biggl[ p^2 \bigl( 1+6 \alpha \phi + \\
         & \quad 8 \alpha ^2 \phi^2 \bigr) - p \Bigl\{ 1+ 2\lambda \phi +16\alpha ^2 \phi ^2 \bigl( -1+\lambda \phi \bigr) +4 \alpha \phi \bigl( -1+ 3 \lambda \phi \bigr) \Bigr\} + \phi^2 \Bigl\{ \lambda^2 + 2\alpha \lambda \bigl( -5 +3 \lambda \phi \bigr) + \\
         & \quad 8 \alpha^2 \bigl( 1- 3 \lambda \phi + \lambda^2 \phi^2 \bigr) \Bigr\} \biggr] + 16 e^{\lambda \phi} \beta \phi^{p+1} \biggl\{ 2p^2 \bigl( 3+ 8 \alpha \phi \bigr)+ p \Bigl\{ 2 - 12 \lambda \phi -4 \alpha \phi \bigl( -5 + 8\lambda \phi \bigr) \Bigr\} + \phi \\
         & \quad \Bigl\{ \lambda \bigl( -5 + 6\lambda \phi \bigr) + 2 \alpha \bigl( 3 -14 \lambda \phi +8 \lambda^2 \phi^2 \bigr)  \Bigr\} \biggr\} V_0 +256 \beta^2 \phi^{2+2p} \biggl\{ 1+6p^2 -7 \lambda \phi +6 \lambda^2 \phi^2 + p \bigl( 5-\\
         & \quad 12 \lambda \phi \bigr) \biggr\} V_0^2\Biggr]\\
    \end{split}
\end{equation}
\noindent The CMBR spectral index parameters $n_s$  and $r$ will be modified accordingly too,
\begin{equation}
    \begin{split}
       n_s
       &=1 - \frac{3 e^{\lambda  \phi } \biggl\{e^{\lambda  \phi } \Bigl\{\phi  \bigl(4 \alpha  \lambda  \phi -4 \alpha +\lambda \bigr)-p \bigl(1 +4 \alpha  \phi \bigr) \Bigr\}+16 \beta V_0 \phi ^{p+1} \bigl(2 \lambda  \phi -2 p-1 \bigr) \biggr\}^2}{ \Bigl\{ \bigl( 1 + 2 \alpha  \phi \bigr) e^{\lambda  \phi }+16 \beta  V_0 \phi ^{p+1}\Bigr\} \Bigl\{\phi \bigl( 1 +4 \alpha  \phi \bigr) e^{\lambda  \phi}+16 \beta V_0 \phi ^{p+2}\Bigr\}^2} + \\
       & \quad \frac{2}{\Bigl\{ \bigl( 1+ 4 \alpha  \phi  \bigr) e^{\lambda  \phi }+16 \beta  V_0 \phi ^{p+1}\Bigr\} \Bigl\{\phi \bigl( 1+ 2 \alpha  \phi \bigr) e^{\lambda \phi }+16 \beta V_0 \phi^{p+2} \Bigr\}^2} \times  e^{\lambda \phi} \Biggl[ e^{2 \lambda  \phi } \biggl[ p^2 \bigl( 1+6 \alpha \phi + \\
         & \quad 8 \alpha ^2 \phi^2 \bigr) - p \Bigl\{ 1+ 2\lambda \phi +16\alpha ^2 \phi ^2 \bigl( -1+\lambda \phi \bigr) +4 \alpha \phi \bigl( -1+ 3 \lambda \phi \bigr) \Bigr\} + \phi^2 \Bigl\{ \lambda^2 + 2\alpha \lambda \bigl( -5 +3 \lambda \phi \bigr) + \\
         & \quad 8 \alpha^2 \bigl( 1- 3 \lambda \phi + \lambda^2 \phi^2 \bigr) \Bigr\} \biggr] + 16 e^{\lambda \phi} \beta \phi^{p+1} \biggl\{ 2p^2 \bigl( 3+ 8 \alpha \phi \bigr)+ p \Bigl\{ 2 - 12 \lambda \phi -4 \alpha \phi \bigl( -5 + 8\lambda \phi \bigr) \Bigr\} + \phi \\
         & \quad \Bigl\{ \lambda \bigl( -5 + 6\lambda \phi \bigr) + 2 \alpha \bigl( 3 -14 \lambda \phi +8 \lambda^2 \phi^2 \bigr)  \Bigr\} \biggr\} V_0 +256 \beta^2 \phi^{2+2p} \biggl\{ 1+6p^2 -7 \lambda \phi +6 \lambda^2 \phi^2 + p \bigl( 5-\\
         & \quad 12 \lambda \phi \bigr) \biggr\} V_0^2\Biggr]\\
    \end{split}
\end{equation}
and 
\begin{equation}
    \begin{split}
        r =\frac{8 e^{\lambda  \phi } \biggl\{e^{\lambda  \phi } \Bigl\{\phi  \bigl(4 \alpha  \lambda  \phi -4 \alpha +\lambda \bigr)-p \bigl(1 +4 \alpha  \phi \bigr) \Bigr\}+16 \beta V_0 \phi ^{p+1} \bigl(2 \lambda  \phi -2 p-1 \bigr) \biggr\}^2}{ \Bigl\{ \bigl( 1 + 2 \alpha  \phi \bigr) e^{\lambda  \phi }+16 \beta  V_0 \phi ^{p+1}\Bigr\} \Bigl\{\phi \bigl( 1 +4 \alpha  \phi \bigr) e^{\lambda  \phi}+16 \beta V_0 \phi ^{p+2}\Bigr\}^2}
    \end{split}
\end{equation}
Here, we have considered a fixed value of $\beta = 10^{-6}$ and found the range of $\alpha$ for each potential. The range of $\alpha$ for $p=2,\lambda=0,0.01,0.1$ are found to be $[1\times 10^{-3}, 0.025]$, $[5.9\times 10^{-3},0.01709]$ and $[2.3\times 10^{-3},0.0252]$ whereas for $p=4,\lambda=0,0.01,0.1$ the values of $\alpha$ are found to lie within $[0.15,0.32]$, $[0.325,0.429]$, $[0.7876,0.9469]$. These ranges are obtained from the requirement of compatibility of spectral index parameters with the observational data and they are shown in Table~(\ref{table:3a}). 
\begin{table}[h]
\centering
\addtolength{\tabcolsep}{1.5pt}
\small
\begin{tabular}{cccccccccc}
\hline
Potential, & $V=V_0\phi^p e^{-\lambda\phi}$, & $p = 2$ & & & & & & & \\

Range of $\alpha$ & $\alpha$ & $\beta$ & $\lambda$ & $\phi $ &  $\phi_f$ & N & $n_s$ & r \\ 
\hline
$0.001<\alpha<0.0250$ & 0.02 & $10^{-6}$ & 0 & 15 & 1.44635 & 55 & 0.97629 & 0.14579.\\
$0.00589< \alpha < 0.01709$ & 0.01500 & $10^{-6}$ & 0.01 & 15.22 & 1.42969 & 59 & 0.97671 & 0.13038\\
$0.00232< \alpha < 0.02520$ & 0.01000 & $10^{-6}$ & 0.1 & 11.4 & 1.33780 & 46 & 0.96803 & 0.06951\\
\hline
Potential, & $V=V_0\phi^p e^{-\lambda\phi}$, & $p = 4$ & & & & & & & \\

Range of $\alpha$ & $\alpha$ & $\beta$ & $\lambda$ & $\phi $ &  $\phi_f$ & N & $n_s$ & r \\ 
\hline
 $0.1500<\alpha<0.3200$ & 0.25 & $10^{-6}$ & 0 & 12.5 & 2.27138 & 58 & 0.98082 & 0.16995. \\
$0.32470< \alpha < 0.42900$ & 0.38000 & $10^{-6}$ & 0.01 & 9.5 & 2.08463 & 37 & 0.97571 & 0.24767\\
$0.78760< \alpha < 0.94690$ & 0.90000 & $10^{-6}$ & 0.1 & 7.5 & 1.66064 & 38 & 0.97641 & 0.17378\\
\hline
\end{tabular}
\caption {\label{table:3a} For $V=V_0\phi^{p}e^{-\lambda\phi}$, the e-fold number $N$ and the spectral index parameters $n_s$, $r$ and $n_T$, calculated for a fixed value of $\phi$,$\alpha$ and $\beta$ are presented.}
\end{table}

\subsubsection{Inflaton Potential $V=V_0\frac{\lambda\phi^p}{1+\lambda\phi^p}$}
\begin{figure}[htp]
 \centerline {\psfig{file=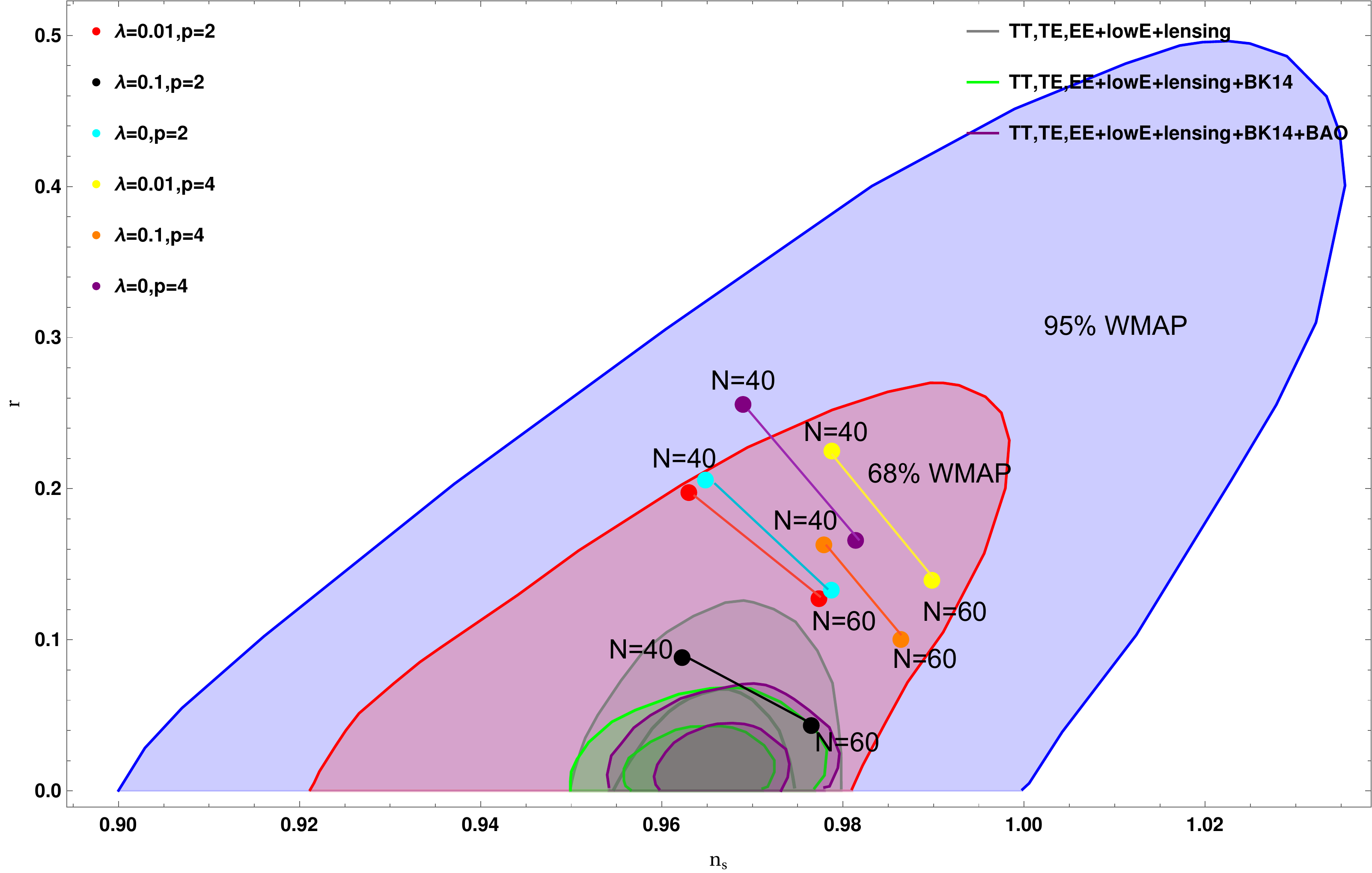,width=6.5cm}\psfig{file=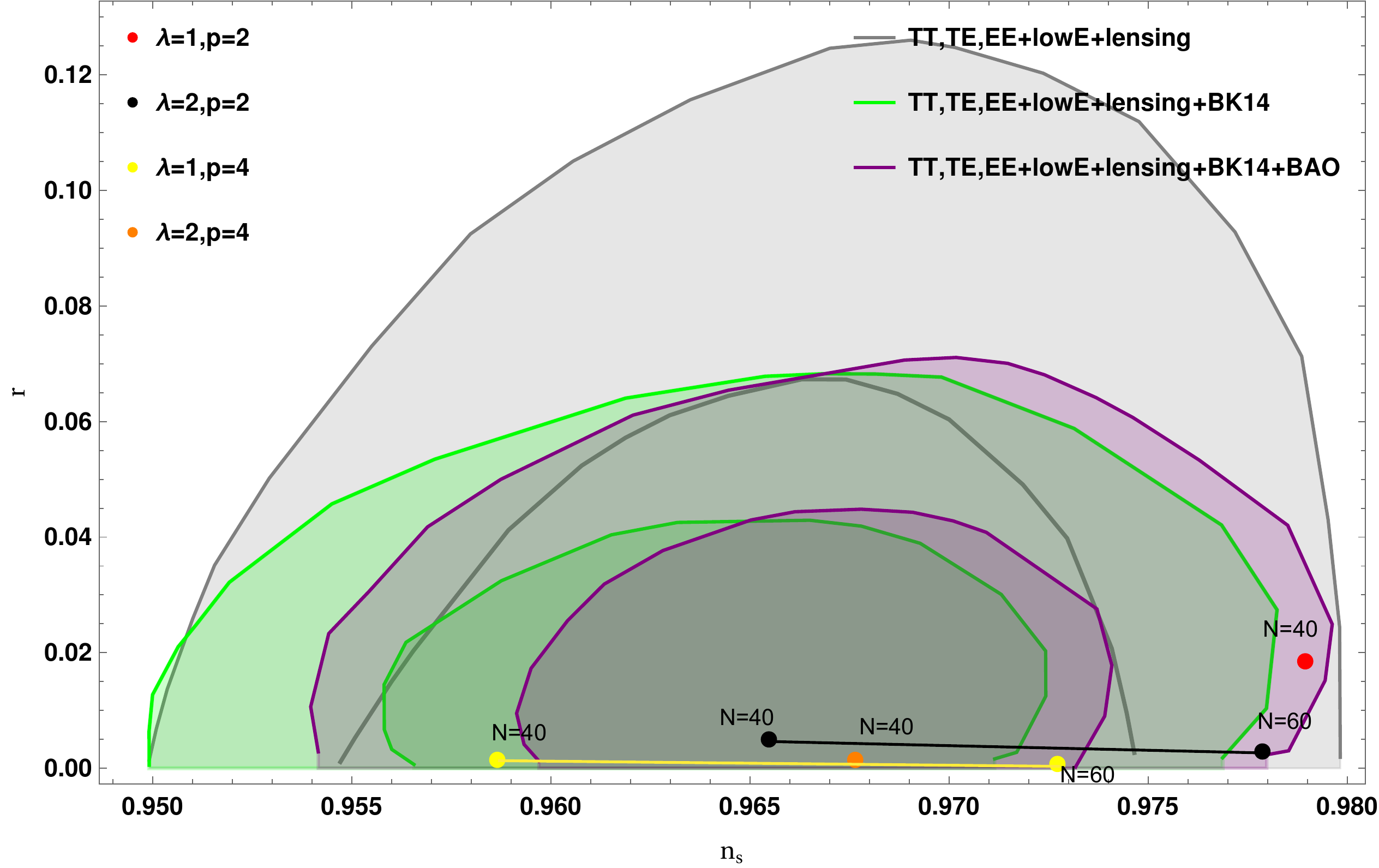,width=6.5cm}}
 \vspace*{3pt}
  \caption{(Color online) Constraints on $n_s$ and $r$ from CMB measurements of different potential. Shaded regions are allowed by WMAP measuremnts, PLANCK alone, PLANCK+BK15, PLANCK+BK15+BAO upto $68\%$ and $95\%$ Confidence Level. \label{Plot3}}
\end{figure}

With the $T$ and $T^2$ terms in $f(\phi,T)$, the slow-roll parameters are derived as 
\begin{equation}
    \begin{split}
        \epsilon_V = \frac{\Bigl\{p+p\lambda\phi^p \bigl(1+32V_0\beta\phi \bigr)+4p\alpha\phi \bigl(1+\lambda\phi^p \bigr)+4\phi \bigl(1+\lambda\phi^p \bigr)\bigl(\alpha+\alpha\lambda\phi^p+4V_0\beta\lambda\phi^p \bigr) \Bigr\}^2}{2\phi^2 \bigl(1+\lambda\phi^p \bigr)\Bigl\{1+\lambda\phi^p \bigl(1+16V_0\beta\phi \bigr)+2\alpha\phi\bigl(1+\lambda\phi^p \bigr)\Bigr\}\Bigl\{1+\lambda\phi^p \bigl(1+16V_0\beta\phi \bigr)+4\alpha\phi\bigl(1+\lambda\phi^p \bigr)\Bigr\}^2}
    \end{split}
\end{equation}
and 
\begin{equation}
    \begin{split}
        \eta_V
        &= \frac{1}{\phi^2 \bigl(1+\lambda\phi^p \bigr)\Bigl\{1+\lambda\phi^p(1+16V_0\beta\phi)+2\alpha\phi \bigl(1+\lambda\phi^p \bigr)\Bigr\}^2 \Bigl\{1+\lambda\phi^p \bigl(1+16V_0\beta\phi \bigr)+4\alpha\phi \bigl(1+\lambda\phi^p \bigr) \Bigr\}}\times\\
        &\quad \Biggl[8\phi^2 \bigl(1+\lambda\phi^p \bigr)\Bigl\{32V_0\beta^2\lambda^2\phi^{2p}+12V_0\alpha\beta\phi^p \bigl(1+\lambda\phi^p \bigr)+\bigl(\alpha+\alpha\lambda\phi^p \bigr)^2\Bigr\}+p^2 \biggl\{ 1+\lambda\phi^p \bigl(1+96V_0\beta\phi \bigr)-\\
        &\quad \lambda^3\phi^{3p}\bigl(1+48V_0\beta\phi+512V_0\beta^2\phi^2 \bigr)+\lambda^2\phi^{2p}\bigl(-1+48V_0\beta\phi+1536 V_0^2\beta^2\phi^2 \bigr)-8\alpha^2\phi^2 \bigl(-1+\lambda\phi^p \bigr)\times\\
        &\quad \bigl(1+\lambda\phi^p \bigr)^2-2\alpha\phi \bigl(1+\lambda\phi^p \bigr) \Bigl\{-3-128 V_0\beta\lambda\phi^{1+p}+\lambda^2\phi^{2p} \bigl(3+64V_0\beta\phi \bigr)\Bigr\}\biggr\}+p \bigl(1+\lambda\phi^p \bigr)\biggl\{-1+\\&2\lambda\phi^p \bigl(-1+16V_0\beta\phi \bigr)+\lambda^2\phi^{2p}\bigl(-1+32V_0\beta\phi+1280V_0^2\beta^2\phi^2 \bigr)+16\alpha^2\phi^2 \bigl(1+\lambda\phi^p \bigr)^2+4\alpha\phi \bigl(1+\lambda\phi^p \bigr)\times\\
        &\quad \Bigl\{1+\lambda\phi^p \bigl(1+80V_0\beta\phi \bigr) \Bigr\} \biggr\} \Biggr]
    \end{split}
\end{equation}
The scalar spectral index $n_s$ and tensor-to-scalar ration $r$ are calculated as 
\begin{equation}
    \begin{split}
        n_s
        &= \frac{1}{\phi^2 \bigl(1+\lambda\phi^p \bigr)\Bigl\{1+\lambda\phi^p \bigl(1+16V_0\beta\phi \bigr) +2\alpha\phi \bigl(1+\lambda\phi^p \bigr)\Bigr\}^2 \Bigl\{1+\lambda\phi^p \bigl(1+16V_0\beta\phi \bigr)+4\alpha\phi \bigl(1+\lambda\phi^p \bigr)\Bigl\} }\\
        &\quad \Biggl[ 1-\frac{3\Bigl\{p+p\lambda\phi^p \bigl(1+32V_0\beta\phi \bigr)+4p\alpha\phi \bigl(1+\lambda\phi^p \bigr)+4\phi \bigl(1+\lambda\phi^p \bigr)\bigl(\alpha+\alpha\lambda\phi^p+4V_0\beta\lambda\phi^p \bigr)\Bigr\}^2}{\phi^2 \bigl(1+\lambda\phi^p \bigr)\Bigl\{1+\lambda\phi^p \bigl(1+16V_0\beta\phi \bigr)+2\alpha\phi \bigl(1+\lambda\phi^p \bigr)\Bigr\}\Bigr\{1+\lambda\phi^p \bigl(1+16V_0\beta\phi \bigr)+4\alpha\phi \bigl(1+\lambda\phi^p \bigr)\Bigl\}^2}+\\
        &\quad 2\biggl\{8\phi^2(1+\lambda\phi^p)^2 \Bigl\{32V_0^2\beta^2\lambda^2\phi^{2p}+12V_0\alpha\beta\lambda\phi^p \bigl(1+\lambda\phi^p \bigr)+\bigl(\alpha+\alpha\lambda\phi^p \bigr)^2\Bigr\}+p^2 \Bigl\{1+\lambda\phi^p(1+96V_0\beta\phi)-\\
        &\quad \lambda^3\phi^{3p}\bigl(1+48V_0\beta\phi+512V_0^2\beta^2\phi^2 \bigr)+\lambda^2\phi^{2p}\bigl(-1+48V_0\beta\phi+1536V_0^2\beta^2\phi^2 \bigr)-8\alpha^2\phi^2 \bigl(-1+\lambda\phi^p \bigr)\times\\
        &\quad \bigl(1+\lambda\phi^p \bigr)^2-2\alpha\phi \bigl(1+\lambda\phi^p \bigr) \Bigl(-3-128V_0\beta\lambda\phi^{1+p}+\lambda^2\phi^{2p}\bigl(3+64V_0\beta\phi \bigr)\Bigr)\Bigr\}+p\bigl(1+\lambda\phi^p \bigr)\Bigl\{-1+\\
        &\quad 2\lambda\phi^p \bigl(-1+16V_0\beta\phi \bigr)+\lambda^2\phi^{2p}\bigl(-1+32V_0\beta\phi+1280V_0^2\beta^2\phi^2 \bigr)+16\alpha^2\phi^2 \bigl(1+\lambda\phi^p \bigr)\Bigr\}\biggr\}\Biggr]
    \end{split}
\end{equation}
and 
\begin{equation}
    \begin{split}
        r
        &=\frac{8\Bigl\{p+p\lambda\phi^p \bigl(1+32V_0\beta\phi \bigr)+4p\alpha\phi \bigl(1+\lambda\phi^p \bigr)+4\phi \bigl(1+\lambda\phi^p \bigr)\bigl(\alpha+\alpha\lambda\phi^p+4V_0\beta\lambda\phi^p \bigr) \Bigr\}^2}{\phi^2 \bigl(1+\lambda\phi^p \bigr)\Bigl\{1+\lambda\phi^p \bigl(1+16V_0\beta\phi \bigr)+2\alpha\phi\bigl(1+\lambda\phi^p \bigr)\Bigr\}\Bigl\{1+\lambda\phi^p \bigl(1+16V_0\beta\phi \bigr)+4\alpha\phi\bigl(1+\lambda\phi^p \bigr)\Bigr\}^2}
    \end{split}
\end{equation}
In Table~(\ref{table:3b}), for the potential $V=V_0\frac{\lambda\phi^p}{1+\lambda\phi^p}$, we have shown the range of $\alpha$ where the different cosmological parameter values are tabulated for particular value of $\phi$ and $\alpha$(chosen from the range(shown)). The range of $\alpha$ for $p=2,\lambda=1,2$ are $[0.002,0.02]$ and $[0.0001,0.0018]$ whereas for $p=4,\lambda=1,2$  $\alpha$ lies within $[10^{-5},0.001]$, $[10^{-5},0.002]$. We see the effect of both $T$ and $T^2$ terms in $f(\phi,T)$ - for the same choices of $\alpha$ and $\beta$, a large value of $r$ is obtained for the potential $V_0 \phi^p e^{\lambda\phi}$, whereas a smaller value of $r$ for $V_0 \frac{\lambda\phi^p}{1+\lambda\phi^p}$ is obtained. 
\begin{table}[h]
\addtolength{\tabcolsep}{1.5pt}
\centering
\small
\begin{tabular}{ccccccccc}
\hline
Potential, & $V=V_0\frac{\lambda\phi^p}{1+\lambda\phi^p}$, & $p = 2$ & & & & & & \\

 range of $\alpha$ &$\alpha$ & $\beta$ & $\lambda$ & $\phi $ &  $\phi_f$ & N & $n_s$ & r \\ 
\hline
$0.002<\alpha<0.02 $&0.01 & $10^{-5}$ & 1 & 5 & 0.84287 & 43 & 0.98108 & 0.01733\\
$0.0001<\alpha<0.0018 $ &0.001 & $10^{-5}$ & 2 & 4 & 0.70788 & 59 & 0.97760 & 0.00294\\
\hline
Potential, & $V=V_0\frac{\lambda\phi^p}{1+\lambda\phi^p}$, & $p = 4$ & & & & & & \\

 range of $\alpha$&$\alpha$ & $\beta$ & $\lambda$ & $\phi $ &  $\phi_f$ & N & $n_s$ & r \\ 
\hline
 $10^{-5}<\alpha<0.001$& 0.0001 & $10^{-6}$ & 1& 3.2 & 1.11392 & 44 & 0.963258 & 0.001195\\
 $10^{-5}<\alpha<0.002$&0.001 & $10^{-6}$ & 2 & 2.97 & 0.984445 & 43 & 0.971226 & 0.001256\\
\hline
\end{tabular}
\caption {\label{table:3b}For $V=V_0\frac{\lambda\phi^p}{1+\lambda\phi^p}$, the e-fold number $N$ and the spectral index parameters $n_s$, $r$ and $n_T$, calculated for a fixed value of $\phi$ ,$\alpha$ and $\beta$ are presented.}
\end{table}
Finally, in Fig.~(\ref{Plot3}), we have plotted the results of two potentials for
$N = 40$ and $60$.  From the Fig.~(\ref{Plot3}), we also see that in case where both $\alpha \neq 0,~\beta \neq 0$, the potential $V_0\frac{\lambda\phi^p}{1+\lambda\phi^p}$ is better compatible in the $n_s - r$ than the potential $ V_0 \phi^p e^{\lambda\phi}$ for $f(\phi,T)=2 \phi (\alpha T + \beta T^2)$.
\newpage 
 \section{Conclusion}
In this manuscript, we have proposed an extension of the modified gravity as $f(\phi,T) =  2 \phi( \kappa^{1/2} \alpha T + \kappa^{5/2} \beta T^2) $ where the inflaton $\phi$ couples linearly and quadratically to the trace of energy-momentum tensor $T$ of the inflaton matter. Such an extension seems to be interesting as it offers an alternative approach to deal with cosmological problems, including dark energy and dark matter. 
 We have investigated the paradigm of inflationary expansion with a specific form of the modified gravity parameter(as above) $f(\phi,T)$ for two distinct inflaton potentials (a)$V=V_0\phi^p e^{-\lambda \phi}$ and (b)$V=V_0\frac{\lambda\phi^p}{1+\lambda\phi^p}$. We have considered three different cases - Case I: $\alpha=0, \beta\neq 0$, Case II: $\alpha\neq 0, \beta=0$ and Case III: $\alpha\neq 0, \beta\neq 0$. We have derived the slow-roll parameters in each of these two potentials and computed the CMBR parameters i.e. the scalar spectral index $n_s$, the tensor to scalar ratio $r$ in order to study the  inflation. 
 With $\alpha \neq 0$ and $\beta = 0$ (Case I) in $f(\phi,T)$, we found that the values of $n_s$ and $r$obtained are in good agreement with the Planck 2018 data. However, when the $\alpha$ term is turned off, $\beta$ is not (Case II), we obtain quite higher values of the scalar-to-tensor ratio while the $n_s$ values are still in good agreement with the Planck 2018 data (Table~\ref{table:1b}).
In the case where $\alpha \neq 0$ and $\beta \neq 0$, the $r$ values, obtained with the potential $V = V_0 \phi^p e^{-\lambda \phi}$, are found to be still higher(by an order) than the one obtained with the potential $V=V_0\frac{\lambda\phi^p}{1+\lambda\phi^p}$. Note that if the coupling term is set to zero i.e. $\alpha = \beta = 0$, we recover the Einstein's gravity. 
 We also found that both our potentials agree quite well with experimental data in their respective parameter space for the simplest form of the modified gravity parameter $R+f(\phi,T)=R+2\alpha\phi T$. It is also to be noted that $\beta$, the coefficient of $T^2$, is very small; hence its contribution to different cosmological parameters is negligible. Also, the range of $\alpha$ becomes smaller when we add higher order terms of $T$.  We find that the inflationary dynamics and the $n_s-r$ values are very sensitive to the coupling parameters  $\alpha$ and $\beta$ in the modified $(\phi - T)$ gravity theory. Finally, we  conclude that the inflaton potential $V=V_0\frac{\lambda\phi^p}{1+\lambda\phi^p}$ fits best for all three cases and all the cosmological parameters lie within $3\sigma$ range of PLANCK 2018 data even for higher order terms.\\
 


 \section*{Acknowledgement}
Ashmita would like to thank BITS Pilani K K Birla Goa campus for the fellowship support. PS would like to thank Department of Science and Technology, Government of India for INSPIRE fellowship. We thank Kinjal Banerjee and Rudranil Basu for the useful discussions and insightful comments related to this work.

\end{document}